\documentclass[aps,twocolumn,showpacs]{revtex4}

\usepackage{graphicx}
\usepackage{latexsym}
\begin{document}

\title{Nature of phase transition(s) in striped phase of \\
triangular-lattice Ising antiferromagnet
}\author{S. E. Korshunov} \affiliation{L. D. Landau Institute for
Theoretical Physics, Kosygina 2, Moscow 119334, Russia}
\date{\today}

\begin{abstract}

Different scenarios of the fluctuation-induced disordering of the striped
phase which is formed at low temperatures in the triangular-lattice
Ising model with the antiferromagnetic interaction
of nearest and next-to-nearest neighbors are analyzed and compared.
The dominant mechanism of the disordering is related to the formation of
a network of domain walls, which is characterized by an extensive number
of zero modes and has to appear via the first-order phase transition.
In principle, this first-order transition can be preceded by a continuous
one, related to the spontaneous formation of double domain walls
and a partial restoration of the broken symmetry, but the realization of such
a scenario  requires the fulfillment of rather special relations between
the coupling constants.

\end{abstract}


\maketitle

\section{Introduction}

An Ising model can be defined by the Hamiltonian
\begin{equation}                                            \label{HIs}
H=\sum_{({\bf i},{\bf j})}J_{\bf ij}\sigma_{\bf i}\sigma_{\bf j}\,,
\end{equation}
where the fluctuating variables (spins), $\sigma_{\bf j}=\pm 1$,
are defined on the sites ${\bf j}$ of some regular lattice.
In the standard version of the model \cite{Ons,Hout,Wan} the coupling
constants, $J_{\bf ij}$, are assumed to be non-zero only when ${\bf i}$ and
${\bf j}$ are the nearest neighbors of each other, but in a more general
case one can suppose that they depend on the distance between ${\bf i}$
and ${\bf j}$.
The models belonging to this class play an extremely important role in the
condensed matter physics, because they can be used for the description
of a huge variety of systems with a two-fold degeneracy of an order parameter.
The best known examples of such systems are ferromagnets and
antiferromagnets with strong easy axis anisotropy and absorbed monolayers.

The exact solution of the Ising model on a triangular lattice
with the interaction of only nearest neighbors was found
%
in 1950 \cite{Hout,Wan}. In the case of the isotropic antiferromagnetic
interaction it demonstrates rather unusual properties.
Namely, the system  remains disordered at arbitrarily low temperature
\cite{Hout,Wan} and at zero temperature is characterized by an algebraic
decay of the correlation functions \cite{St64,St70} and a finite residual
entropy per site \cite{Wan}.
The ground states of this model can be mapped \cite{BH} onto the states of
a solid-on-solid (SOS) model describing the fluctuations of the (111) facet
of a crystal with a simple cubic lattice and are infinitely
degenerate.

This degeneracy is not related to symmetry
and therefore in a physical situation its removal by the interactions
of more distant neighbors should be taken into account.
If the interaction of second neighbors (characterized by the coupling
constant $J_2$) is included into consideration, for both signs of $J_2$
the degeneracy of the ground states is reduced to a sixfold one \cite{Met}.
In terms of the SOS representation \cite{BH}, the ferromagnetic interaction
of second neighbors ($J_2<0$) corresponds to the positive energy of a step
and, therefore, leads to the stabilization of the flat phase at low enough
temperatures \cite{NHB}.
With the increase of temperature a roughening transition \cite{Weeks}
takes place, which at a higher temperature is followed by another phase
transition related to the dissociation of pairs of dislocations
\cite{NHB,Land}.
Both transitions belong to the Berezinskii-Kosterlitz-Thouless 
universality class.

This scenario follows also from approximate mappings \cite{AP,FSK}
of the considered model onto a six-state clock model and has been confirmed
by numerous Monte-Carlo simulations \cite{Land,FSK,MC}.
As is typical for low-dimensional systems, a mean field analysis \cite{Mek}
leads to the wrong conclusions about the character of the intermediate phase
or the nature of phase transitions.

\begin{figure}[b]
\vspace{0mm}
\begin{center}
\includegraphics[width=30mm]{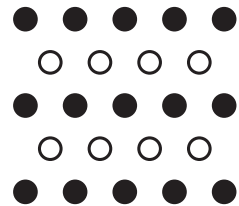}
\end{center}
\caption{\label{1a}
The structure of the ground state for $J_{1,2}>0$.
}\end{figure}

The present work is devoted to the antiferromagnetic Ising model on triangular
lattice in which 
the interaction of second neighbors is also antiferromagnetic.
For brevity we shall call such a system
a triangular-lattice Ising antiferromagnet.
The structure of the ground state of the triangular-lattice Ising
antiferromagnet with the interaction  of first and second neighbors
\cite{Met} is shown in Fig. \ref{1a}.
Here and below we use filled and empty circles
to denote the spins of opposite signs.

Although this version of the Ising model have been also investigated by
different methods  \cite{DSWG,KTK,SH,EH},
its properties are not as clearly understood
as those of the model with the ferromagnetic interaction of second neighbors.
In particular, Domany {\em et al.} \cite{DSWG} have demonstrated
that the formal construction of the Ginzburg-Landau functional describing
the formation of the state 
shown in Fig. \ref{1a} reproduces the Hamiltonian of the Heisenberg
model with the cubic anisotropy, which under the renormalization is
transformed \cite{DR} into the discrete cubic model \cite{KLF}.
However, this hardly allows one to make any conclusions about
the properties of a triangular-lattice Ising antiferromagnet (i) because
the cubic model allows for {\em three} different scenarios of disordering
\cite{NRS}
(although sometimes only one of them is mentioned \cite{Sch}),
and (ii) because this approach does not take into
account the strong chiral asymmetry \cite{OH}
existing in the problem (see Sec. \ref{gs}).
Additionally, such a description cannot reproduce the well-known properties
of the model with only first-neighbor interaction. The same can be also said
about the mean-field approach of Kaburagi {\em et al.} \cite{KTK}, which
gives an unrealistic prediction of a three-sublattice intermediate phase.

A more direct analysis of the fluctuations which can be responsible for
the disordering of the striped phase has been undertaken
by Hemmer {\em et al.} \cite{SH,EH}.
However, these authors have considered
the formation of only one type of domain walls, whose appearance
(as is discussed in Sec. \ref{ddw}) can lead
only to a {\em partial} restoration of the broken symmetry.
Thus, all conclusions of Ref. \onlinecite{SH} and Ref. \onlinecite{EH} are
based on the essentially incomplete physical picture and have to be reconsidered.

This article is devoted to comparison of different mechanisms and different
scenarios of disordering which are possible in the striped phase
of a triangular-lattice Ising antiferromagnet.
Although a number of numerical works \cite{GP,NKL,RRT} give evidence on
the existence of a single first-order transition
(for this or that choice of the relation between the coupling constants),
in a situation when different mechanisms of disordering compete with each
other one hardly can be sure about the universality of this result.

Our analysis assumes that $J_1$ is much larger then all other
coupling constants, which allows us to use the analytical methods based on
the separation of different energy scales.
This limit is also of interest because it corresponds to the case of
strong chiral asymmetry and is the most relevant one
for many physical realizations of the model.
The outlook of the article is as follows.

In Sec. \ref{gs} we discuss the symmetry of the ground states and
the structure of domain walls and estimate the temperature at which
the free energy of a single domain wall vanishes as a result
of thermal fluctuations of this wall.
In Sec. \ref{ddw} we show that the free energy of a double domain wall can
be expected to vanish at much lower temperature than that of a single wall,
and argue that this suggests a possibility of a two-transition scenario.

In Sec. \ref{dwnw} the spontaneous formation of a network of single domain
walls is analyzed. In the limit when this network has to be diluted, it
becomes clear that it has to appear via a first-order phase transition.
When only the interaction of up to third 
neighbors is taken into account,
the estimates for the temperatures of this transition and of
the spontaneous formation of double domain walls coincide with each other,
which suggests that there is only one phase transition in the system.
In Sec. \ref{htph} a phenomenological free-energy functional
is constructed which allows one to confirm that the transition
to the disordered phase has always to be of the first order.

The interplay between the two main mechanisms of the disordering is analyzed
in Sec. \ref{md}, whereas the results are summarized in Sec. \ref{conc}.
The short Appendix is devoted to a formal derivation of an exact upper
boundary for the temperature at which the long-range order in
$\sigma_{\bf j}$ is destroyed by thermal fluctuations when $J_1=\infty$.

\section{Ground states and domain walls\label{gs}}

\begin{figure}[b]
\vspace{0mm}
\begin{center}
\includegraphics[width=35mm]{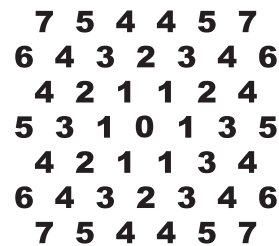}
\end{center}
\caption{\label{1b}
The classification of neighbors on triangular lattice.
}\end{figure}

The structure of the ground state of the triangular-lattice Ising model
with the antiferromagnetic interaction of nearest and second neighbors
\cite{Met} is shown in Fig. \ref{1a},
whereas Fig. \ref{1b} illustrates the classification of neighbors
on a triangular lattice according to their distance from a given cite
(denoted by zero).
In the following we assume that $J_3$, the coupling constant describing
the interaction of third neighbors, can also be non-zero, but
satisfies the constraint \cite{TM,KK}, $J_3<J_2/2$,
which is required for the stability of the striped ground state of
Fig. \ref{1a}. The role of the interactions of more distant neighbors
will be discussed in Sec. \ref{md}.
To make a situation more transparent we assume that the interaction
of nearest neighbors, $J_1$, is much larger then all other coupling constants.

The sixfold degeneracy of the state shown in Fig. \ref{1a}
corresponds to the violation of the $Z_2\times Z_3$ symmetry,
where $Z_2$ is related to the possibility of interchanging positive and
negative spins, and $Z_3$ to three possible orientations of the stripes
formed by spins of the same sign.
Each of the six ground states
can be associated with  a unit vector pointing either in positive or
negative direction along one of the three axes, which are perpendicular to
each other. These six directions can also be put into correspondence with
the six faces of a cube.
In such a representation the group $Z_2$ is related to the reflection
symmetry which transforms the opposite faces of a cube into each other,
whereas the group $Z_3$ corresponds to the cyclic permutations of the three axes.

In systems with a discrete degeneracy the destruction of a long-range
order has to be driven by thermal activation of infinite domain walls.
The appearance of a sequence of more or less parallel walls is expected when
the intersections of walls with
different orientations are energetically unfavored \cite{BMVW}.
The alternative option consists on the appearance of a network of
intersecting domain walls with different orientations, which is would be
favored by a negative energy of domain wall intersections \cite{BMVW}.

\begin{figure}[b]
\vspace{0mm}
\begin{center}
\includegraphics[width=70mm]{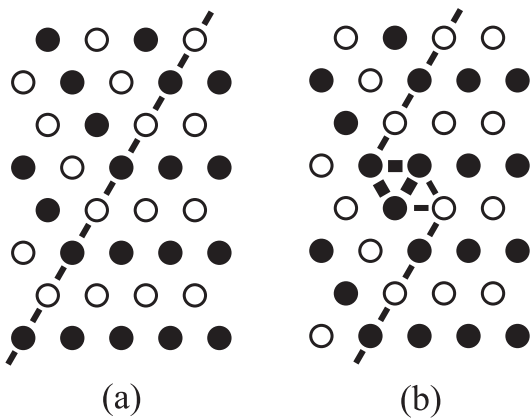}
\end{center}
\caption{\label{2} Low energy domain walls: (a) a straight one;
(b) with a kink.}
\end{figure}

Fig. \ref{2}(a) shows an example of a lowest-energy domain wall separating
two different ground states.
Each segment of such a wall connects two neighboring spins of
the opposite signs and separates two second neighbors of the same sign.
The energy of this wall per unit length,
\[
E_{\rm dw}^{}=2J_2-4J_3>0\,,
\]
does not depend on $J_1$, because its presence does not
lead to the violation of the constraint
\begin{equation}                                          \label{cond}
    \sigma_{\bf j}\sigma_{\bf j'}+\sigma_{\bf j'}\sigma_{\bf j''}+\sigma_{\bf j''}\sigma_{\bf j}=-1
\end{equation}
on any triangular plaquette. This condition is satisfied when
a plaquette contains spins of both signs.

When crossing a domain wall the direction of the stripes formed
by the spins of the same sign changes by $60^{\circ}$  and, therefore,
the direction of a lowest-energy wall is uniquely determined by the
pair of ground states which it separates.
In other terms, the energy of a domain wall in the considered model is strongly dependent both
on its orientation and on which states it separates.
From the analysis of systems with a threefold degeneracy it is known
that such a property (the chiral asymmetry \cite{OH}) can lead to the change
of the universality class \cite{HF} or even of the order
\cite{Vil80a,Vil80b,CFHLB} of a phase transition.
This is the reason why the analogy with the cubic model
\cite{KLF,NRS} without chiral assymetry (which follows from the
Landau-Ginzburg analysis of Domany {\em et al.} \cite{DSWG})
is insufficient for understanding the properties of a triangular-lattice
Ising antiferromagnet with $J_{2,3}\ll J_1$.

Any fluctuations of domain wall
are impossible without the violation of the constraint (\ref{cond}), and
therefore require the energies proportional to $J_1\gg E_{\rm dw}$.
Figure \ref{2}(b) shows an example of the simplest elementary defect (a kink)
which can be formed on a straight domain wall. The energy of such a
defect,
\[
E_{\rm k}=2J_1-4J_3\,,
\]
has to include a contribution proportional
to $J_1$, because the formation of a kink requires to have one
plaquette at which all three spins are of the same sign,
and therefore
\[
   \sigma_{\bf j}\sigma_{\bf j'}+\sigma_{\bf j'}\sigma_{\bf j''}
   +\sigma_{\bf j''}\sigma_{\bf j}=3\,.
\]
In Fig. \ref{2}(b) this plaquette is shown by bold lines.

At finite temperatures the free energy of a domain wall (per unit length)
can be estimated as the difference between its energy and the entropic
term related to the formation of kinks \cite{PD}.
For $T\ll E_{\rm k}$ this gives
\begin{equation}                                          \label{Fdw}
F_{\rm dw}(T)\approx E_{\rm dw}-2T\exp(-E_{\rm k}/T)\,.
\end{equation}

It is well known that when the intersections of domain walls are unfavored,
one can expect the formation of a dilute sequence of parallel
(on the average) walls when the free energy
of a single wall, $F_{\rm dw}(T)$, becomes equal to zero \cite{BMVW,PT}.
For $F_{\rm dw}(T)$ defined by Eq. (\ref{Fdw}) this takes place at
\begin{equation}                                           \label{Tdw}
T=T_1\approx \frac{E_{\rm k}}{\ln(E_{\rm k}/E_{\rm dw})}
\end{equation}
where we have taken into account that $E_{\rm dw}\ll E_{\rm k}$ (as a
consequence of $J_2\ll J_1$).
Eq. (\ref{Tdw}) shows that for \makebox{$J_{2},J_3\ll J_1$} the temperature $T_1$
only weakly depends on $J_2$ and $J_3$ and is determined mainly
by $J_1$.

In the limit $J_1\rightarrow\infty$ the fluctuation-induced vanishing
of the free energy of a single domain wall becomes impossible.
This manifests itself in the divergence of the expression for $T_1$
given by Eq. (\ref{Tdw}) for $E_{\rm k}\rightarrow\infty$.
Quite remarkably, even in this limit there still
remain possibilities for a fluctuation-induced destruction of the
long-range order.
They are related to the spontaneous formation of {\em double} domain walls
(which is discussed in Sec. \ref{ddw})
and of a domain-wall network (discussed in Sec. \ref{dwnw}).

\section{Spontaneous formation of double domain walls\label{ddw}}

\begin{figure}[b]
\vspace{0mm}
\begin{center}
\includegraphics[width=70mm]{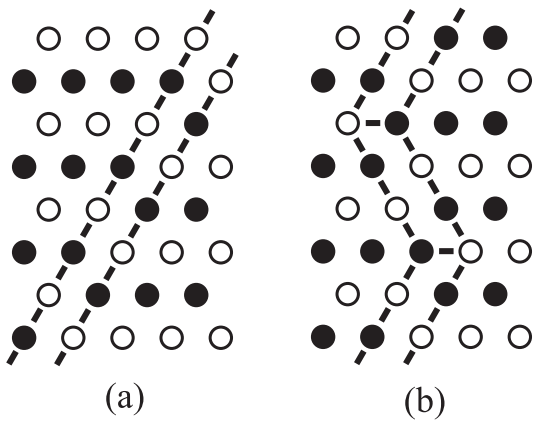}
\end{center}
\caption{\label{3} Double domain walls: (a) straight;
(b) with two corners.}
\end{figure}

A double domain wall consists of two parallel single walls
[see Fig. \ref{3}(a)] and separates two ground states with
the same direction of spin stripes.
The energy of a straight double wall  per segment,
$E_{\rm ddw}^{}$, is given simply by the energy of two single domain walls
from which it consists, $E_{\rm ddw}=2E_{\rm dw}$. The interaction between
two parallel single walls does not appear even if one takes into account
the interactions of spins with their fourth and fifth
neighbors (see Fig. \ref{1b}). 
However, it turns out that the fluctuations of a double wall cost less energy
than those of a single wall, as a consequence of which its
free energy can easily become smaller than that of a single wall.

Fig. \ref{3}(b) demonstrates that the fluctuations of a double domain wall
are possible without the violation of the constraint (\ref{cond}).
From this figure it is clear that each corner on a double wall
requires the appearance of an additional segment of a single domain wall.
The energy of such a defect, $E_{\rm c}=2J_2$, does not depend on $J_1$ and $J_3$.
At finite temperatures the expression for the free energy of a double wall
which takes into account the presence of thermally
activated corners is of the form \cite{SW,EH}
\begin{equation}                                      \label{Fddw}
F_{\rm ddw}^{}(T)
=2E_{\rm dw}-T\ln[1+\exp(-E_{\rm c}/T)]\,.
\end{equation}
The spontaneous appearance of a diluted sequence of such walls can be
expected to take place when $F_{\rm ddw}^{}(T)$ becomes equal to zero.
The condition $F_{\rm ddw}^{}(T_{\rm })=0$ can be rewritten as
\begin{equation}                                      \label{Fddw=0}
T_{\rm }=\frac{E_{\rm c}}{-\ln[\exp(2E_{\rm dw}/T_{\rm })-1]^{}}\;.
\end{equation}
Note that Eq. (\ref{Fddw=0}) and, therefore its solution,
$T_2$, do not depend on $J_1$.

In the case of \makebox{$J_3=0$} the solution of Eq. (\ref{Fddw=0})
gives
\begin{equation}                                      \label{Tddw0}
T_2= \gamma_2 J_2
\end{equation}
where \cite{SW,EH}
\[
\gamma_2=\frac{1}{2} \ln\!\left[\!
\left(\frac{1}{2}+\sqrt{\frac{23}{108}}\right)^{1/3}\hspace*{-4mm}+
\left(\frac{1}{2}-\sqrt{\frac{23}{108}}\right)^{1/3}\right]
\approx 7.112                      
\]
whereas for \makebox{$E_{\rm dw}\ll E_{\rm c}$} an expansion
of the exponent in the right-hand side of Eq. (\ref{Fddw=0})
allows one to find that
\begin{equation}                                      \label{Tddw}
T_2
\approx \frac{E_{\rm c}}{\ln{(E_{\rm c}/2 E_{\rm dw})}}
\approx \frac{2J_2}{\ln{(J_{\rm 2}/2 E_{\rm dw})}}\;.
\end{equation}

Comparison of Eq.(\ref{Tddw0}) and Eq. (\ref{Tddw}) with Eq. (\ref{Tdw})
demonstrates that for $J_2\ll J_1$ the destruction of the long-range order in
$\sigma_{\bf j}$ cannot be driven by the spontaneous formation of a sequence of
single domain walls, because the analogous sequence of double domain walls
can be expected to appear at much lower temperature.
A numerical calculation of $T_2$ for an arbitrary relation between $J_1$
and $J_2$ (at $J_3=0$) in terms of the one-dimensional SOS model which
takes into account also more complex fluctuations of a double domain wall
can be found in \makebox{Ref. \onlinecite{SH}.}

\begin{figure}[t]
\vspace{3mm}
\begin{center}
\includegraphics[width=65mm]{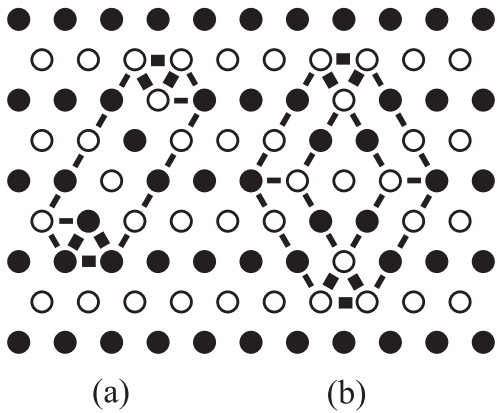}
\end{center}
\caption{\label{loops} Closed loops formed by (a) single domain walls,
(b) double domain walls.}
\end{figure}

In the limit $J_1\rightarrow\infty$ at low temperatures 
the system has to be completely frozen in one of its
ground states,
because any finite size fluctuation on the background of
the striped ground  state requires the violation of the constraint
(\ref{cond}) on at least two plaquettes (see Fig. \ref{loops}).
Accordingly, the exact expression for the free energy of a double domain
wall cannot contain any additional contributions related
to the suppression of finite size fluctuations.
This has allowed Shi and Wortis \cite{SW} to conjecture that
in the limit $J_1\rightarrow\infty$ the condition
$F_{\rm ddw}(T)=0$ [with $F_{\rm ddw}$ given by Eq. (\ref{Fddw})] determines
the {\em exact} value of the transition temperature.
However, this conclusion can be valid only if the spontaneous formation of
a diluted sequence of double domain walls is not preceded by the spontaneous
formation of a network of single domain walls (see Sec. \ref{dwnw}).

It is rather evident that the average direction of a fluctuating double wall
containing thermally activated corners will be perpendicular to the direction
of spin stripes on both its sides.
The spontaneous formation of a sequence of such walls
leads to the restoration
of the $Z_2$ symmetry between the two states with the same direction
of stripes and the reduction of the broken symmetry to $Z_3$.

Since the concentration of walls, $\nu(T)$, is restricted by their collisions
(which are responsible for the reduction of the entropy of their
fluctuations),  Shi and Wortis \cite{SW} have concluded that this phase
transition is continuous and belongs to the Pokrovsky-Talapov \cite{PT}
universality class. Accordingly, in the vicinity of $T_2$
one should have $\nu(T)\propto (T-T_2)^{1/2}$.
Note that 
the value of the correlation length describing
the decay of the correlation function
$\langle \sigma_{\bf i}\sigma_{\bf j}\rangle$ in the direction
along the stripes is inversely proportional to $\nu(T)$.

On the other hand, Einevoll and Hemmer \cite{EH} have argued that this phase
transition cannot be continuous, because the temperature at which
$F_{\rm ddw}(T)$ vanishes is different for different directions of a
double wall. In our opinion this conclusion is completely unjustified.
The dependence of $F_{\rm ddw}$ on the direction of the wall manifests
itself in the spontaneous formation of a sequence of walls with the same
(on the average) direction, and is a {\em necessary condition} for the
applicability of the Pokrovsky-Talapov theory \cite{PT} rather than
an obstacle for its validity.

If one assumes that the value of $J_1$ is finite, but large in comparison
with $J_2$, the fluctuations in the low temperature phase become
possible. Each point where a pair of single walls is created
(or annihilates) cost the energy close to $2J_1$, so at low temperatures
there will exist a finite concentration of highly anisotropic loops formed
by such walls [see Fig. \ref{loops}(a)]. However, at $T\approx T_2\ll J_1$
the average distance between them will be much smaller then their size,
as a consequence of which their presence can be neglected.
On the other hand, the size of the closed
loops formed by double domain walls [see Fig. \ref{loops}(b)]
diverges when $T\rightarrow T_2$.
From the theory of the commensurate-incommensurate
transition it is known \cite{BPT} that this is accompanied by the
change of the type of the phase transition
from the Pokrovsky-Talapov universality class \cite{PT} to that of
the Ising  model. However, the behavior will be changed only in a
narrow region around $T_2$, which will be exponentially small in $J_1/T_2$.

Since the considered phase transition is related to the restoration of
$Z_2$ symmetry, the emergence of the Ising critical behavior  looks
rather natural. The different universality class in the case of $J_1=\infty$
can be explained by the extremely anisotropic nature of domain walls in
that limit, which prevents the merging of different domains of the same state.

If the spontaneous formation of a sequence of double domain walls indeed
takes place as a separate phase transition it has to be followed
(with a further increase of temperature) by a second phase transition
related to the restoration of $Z_3$ symmetry.
The completely disordered phase above this transition will look like a
mixture of finite domains of all six ground states.
Since double walls do not change the orientation of stripes,
the second phase transition requires the formation of single domain walls
of all possible orientations.
Above we have found that the spontaneous formation of double domain walls
takes place when the free energy of a single wall is still much larger
then temperature, so it looks rather plausible that the two phase transition
may be well separated from each other.
However, to check if it is really so, the formation of a
network of single domain walls has to be studied in more detail.

\section{Spontaneous formation of a domain-wall network\label{dwnw}}

\begin{figure}[b]
\vspace{3mm}
\begin{center}
\includegraphics[width=65mm]{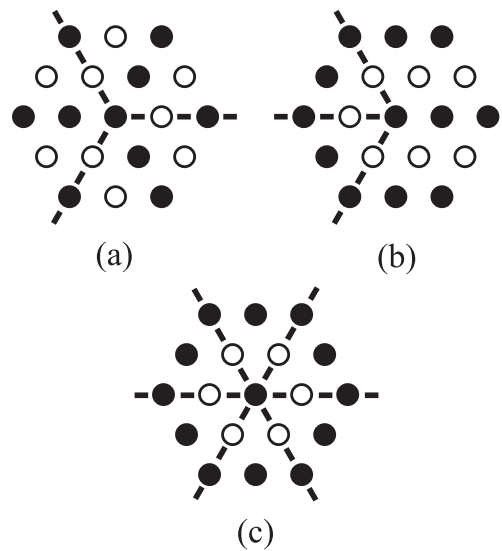}
\end{center}
\caption{\label{5} Three low-energy domain walls with
different orientations can merge or intersect with each other.}
\end{figure}

Like in the previous section it will be convenient to start
by considering the case of $J_1=\infty$.
In this limit all single domain walls have to be straight due to
the absence of kinks.

Fig. \ref{5} shows how such walls can intersect or merge with each other
without violating the constraint (\ref{cond}).
The energy of these intersections does not depend on $J_1$ or $J_2$.
In particular, the energy of the $120^{\circ}$ junction shown
in Fig. \ref{5}(a) is simply equal to zero, $E_{\rm a}=0$,
whereas for the $60^{\circ}$ junction shown in Fig. \ref{5}(b)
it is given by $E_{\rm b}=4J_3$.
The  intersection shown in Fig. \ref{5}(c) can be considered as an
overlap of  two $60^{\circ}$  junctions.
The energy of this object is equal to $E_{*}=12J_3$.

\begin{figure}[b]
\begin{center}
\includegraphics[width=80mm]{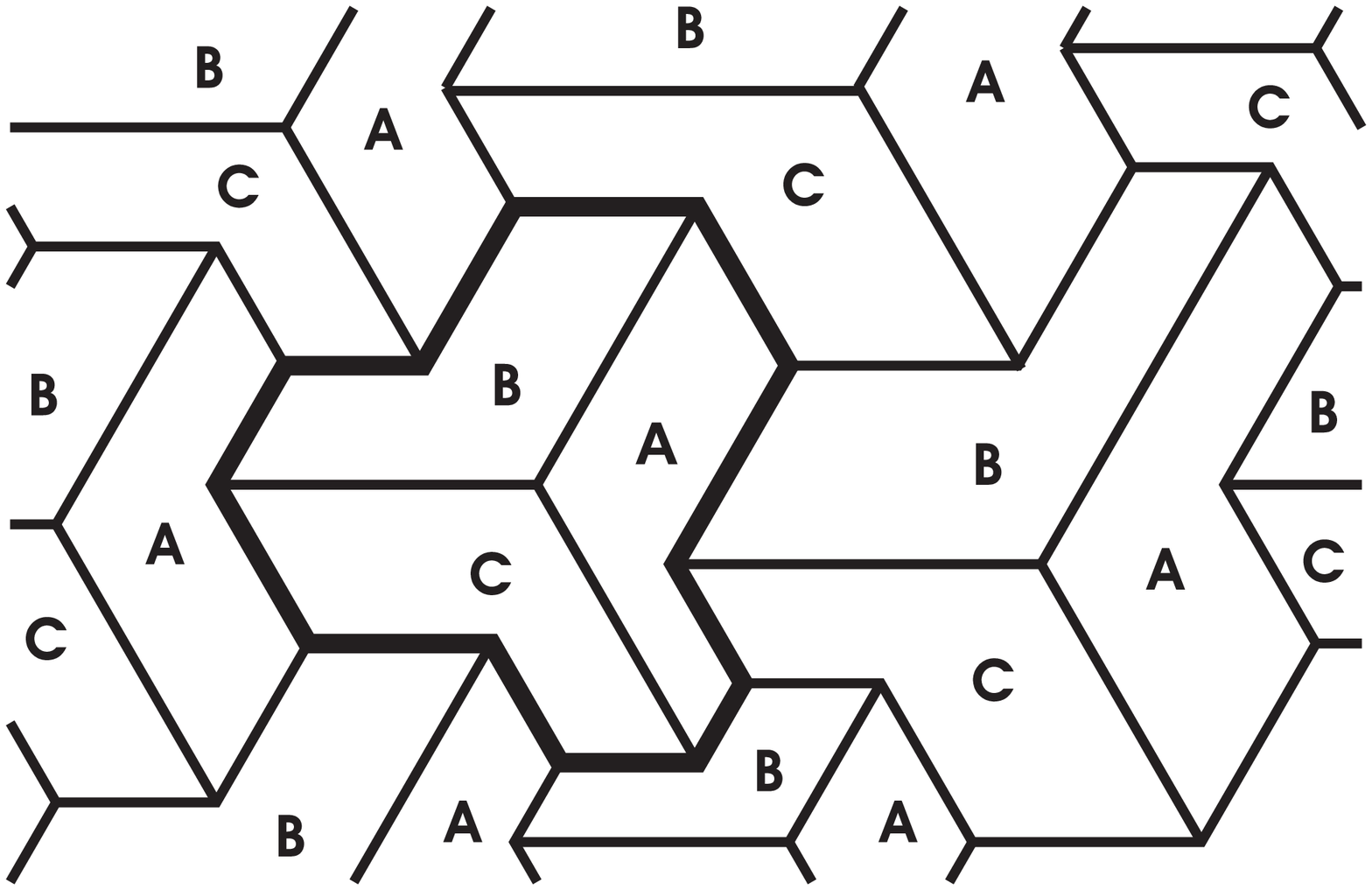}
\end{center}
\vspace{3mm}
\caption{\label{4} A possible structure of a domain-wall network}
\end{figure}

An important feature (already mentioned in Sec. \ref{gs}), which plays
a crucial role in determining the structure of a domain-wall network
for $J_1=\infty$, is that the direction of each wall is uniquely
determined by the states which it separates.
A possible structure of a network which is formed by straight walls
and satisfies this criterion is schematically shown in Fig. \ref{4}.
Here the letters A, B and C are used to denote the domains with three
different orientations of stripes.
Note that all walls between A and B are parallel to each other.
The same is true for all walls between B and C,
         as well as for all walls between C and A.

For the sake of clearness we have not shown in Fig. \ref{4} which of
the two versions of A, of B, or of C (related to the change of sign
of all spins) is realized in each particular domain.
This depends on the exact positions of domain walls.

The structure of the network shown in Fig. \ref{4} has been chosen to
maximize the entropy for the given total length of the walls.
The network of such a kind is characterized by a large number of zero
modes which do not change its energy.
For example, each domain of the type A can be moved to the left or to the
right. This changes the areas of all domains of the types B and C
which are adjacent to it, but the total length of domain walls (and,
therefore, the total energy of the network) is conserved.
Analogously, all domains of the types B and C can be moved in the two other
directions. When a domain is moved by one lattice unit in such a way,
the signs of all spins inside it are reversed.

A combination of all three types of zero modes allows to change the size
of a three-domain complex (a bubble) formed by
neighboring domains of three different types without changing its position.
An example of such a bubble is shown in Fig. \ref{4} by the bold line.
The zero modes of this particular type are called the breathing modes
\cite{CFHLB}.

The existence of breathing modes has been discovered by Villain
\cite{Vil80a} when studying the formation of a honeycomb network
in which each domain has the shape of a hexagon.
In such a network the size of each hexagon can be changed without
changing the total length of domain walls.
A honeycomb network is formed in a system with a threefold degeneracy
in which a domain wall of a given type
(for example, a wall between A and B) can have only
three particular orientations out of six that are generally allowed.
In these terms, in our problem a wall of each type can have only two
orientations out of six, whereas in the three-state Potts model on
a triangular lattice all six orientations are allowed for domain walls of
any type.

The entropy which can be associated with the existence of zero modes can be
estimated as $\ln M$ per mode, where $M$ is the typical number of the
configurations which can be spanned by a given zero mode.
It is clear that in a diluted network $M$ has to be proportional to $L$,
the typical distance between the centers of neighboring bubbles
(in lattice units). This allows one to estimate the free energy
(per unit area) of a network shown in Fig. \ref{4} as
\begin{equation}                                          \label{Fnw}
    F_{\rm nw}(L)\approx \frac{2}{\sqrt{3}}\left[4\frac{E_{\rm dw}}{L}+3\frac{E_{\rm mp}}{L^2}
    -3T\frac{\ln L}{L^2}+O\left(\frac{1}{L^3}\right)\right]\,,
\end{equation}
where
\[
E_{\rm mp}= E_{\rm a}+E_{\rm b}=4J_3\,.
\]
The first term in the right-hand side of Eq. (\ref{Fnw}) describes
the energy of domain walls, the second term - the energy of merging points
and the third term is related to the entropy of zero modes.
The expression (\ref{Fnw}) has the same structure as the one proposed by
Villain \cite{Vil80a,Vil80b} for a honeycomb network.

The variation of $F_{\rm nw}(L)$ with respect to $L$ reveals that
this function has two minima, one of which is situated at $L=\infty$ and
corresponds to the absence of any network and another at $L=L_0<\infty$.
The free energies of these two minima become equal to each other
when 
\begin{equation}                                         \label{L}
    L_0=\frac{3T}{4E_{\rm dw}}\;,~~~\ln L_0=1+\frac{E_{\rm mp}}{T}\;.
\end{equation}
This shows that the applicability of Eq. (\ref{Fnw}), which assumes
$L\gg 1$, requires to have $E_{\rm dw}\ll T\ll E_{\rm mp}$, that is
$E_{\rm dw}\ll J_2$.
With the increase of $E_{\rm dw}/J_2$ the value of $L_0$ at the transition
is decreased and for $E_{\rm dw}\sim J_2$ it becomes comparable with 1, which
means that Eq. (\ref{Fnw}) is no longer applicable.

In the limit of $E_{\rm dw}\ll J_2$ (when $E_{\rm mp}\approx 2J_2$)
the temperature of the first-order
phase transition related to the formation of a domain-wall network, which
follows from Eqs. (\ref{L}), can be estimated as
\begin{equation}                                         \label{Tnw}
    T_{\rm nw}\approx \frac{E_{\rm mp}}{\ln(E_{\rm mp}/E_{\rm dw})}
    \approx\frac{2J_2}{\ln(J_2/E_{\rm dw})}\;,
\end{equation}
whereas the value of $L_0$ (which determines the correlation radius for the
fluctuations of $\sigma_{\bf j}$) at the transition point is given by
\begin{equation}                                         \label{Lc}
L_{c} 
   \approx \frac{3}{2}\frac{J_2/E_{\rm dw}}{\ln(J_2/E_{\rm dw})}\,.
\end{equation}

With the decrease of $J_3$ the ratio $E_{\rm mp}/E_{\rm dw}$ is decreased, which
leads to the decrease of $L_c$. For $J_3\sim J_2$ (that is $E_{\rm dw}\sim E_{\rm mp}$)
the value of $L_c$ following from Eqs. (\ref{L}) becomes comparable
with 1, which means that the approach based on the minimization of
$F_{\rm nw}(L)$ is no longer applicable. However, since the decrease of
$J_3$ makes the first-order nature of the transition more and more
pronounced, one can expect that it will remain of the first order even when
the formation of a domain-wall network does not
allow for a quantitative description.
In the next section this conclusion is confirmed with the help
of a phenomenological analysis which does not take
into account any details of a domain-wall network structure
and, therefore, is applicable in a wide interval of the values of $J_3$
(including $J_3=0$).

The finiteness of $J_1$ cannot be expected to be
of any importance for the phase transition related with the spontaneous
formation of a domain-wall network. It allows for fluctuations of
single domain walls, which no longer have to be straight, but, as has
been shown by Villain \cite{Vil80b} for a hexagonal network, this
does not lead to any qualitative changes.

\section{High-temperature phenomenology\label{htph}}

Like above, it will be convenient to start the analysis
by considering the case of $J_1=\infty$.
In this limit the manifold of the allowed states coincides with the
manifold of the ground states of the system with only nearest neighbor
interaction,
which can be put into correspondence with the states of the (111) facet
of a crystal with a simple cubic lattice \cite{BH} .

In particular, the ground states whose structure is shown in Fig.
\ref{1a} map onto the states with the maximal possible slope, corresponding
to the transformation of the (111) facet into one of the three facets of
the (100) family. The direction of the slope is
determined  by the direction of stripes in spin representation.
These  states do not allow for the formation of
any finite size defects, because this would require locally the further
increase of a slope.

In the limit of $T\rightarrow\infty$ all terms in the partition function
become equal to each other,  therefore in this limit all correlation
functions in the system with $J_1=\infty$ have exactly the same form as
in the model with the interaction of only nearest neighbors at zero
temperature.
The corresponding phase is characterized by the zero slope and
a logarithmical divergence of fluctuations \cite{BH} of the discrete
variable $n$ describing the position of a surface. One can expect
that the same phase will be also stable at large but finite $T$.

In this phase the large-scale fluctuations
of $n$ can be described by a continuous free energy functional,
\begin{equation}                                            \label{f2}
F_{\rm eff}\{n\}=\int d^2{\bf r} f_2\{n\}\,,~~
f_2\{n\}=\frac{K}{2}(\nabla n)^2\;,
\end{equation}
in which the discreteness of $n$ is neglected \cite{BH}.
At $T=\infty$ the dimensionless effective rigidity $K$ (which is of
entropic origin) is equal to $K_0=\pi/9$.
In terms of the SOS representation the energy of a step (per unit length)
is equal to $-2J_2$, therefore for $J_2<0$ the decrease of
temperature is accompanied by a monotonic growth of $K$ \cite{NHB},
which at $K=\pi/2$ leads to the phase
transition to the smooth phase.

We are now considering the opposite case of $J_2>0$, when the decrease
of $T$ from \makebox{$T=\infty$} should be accompanied by the decrease of
$K$ from $K=K_0$.
Since we know that at lower temperatures the triply degenerate
phase with a finite slope has to be formed, the phenomenological
functional (\ref{f2}) has to be replaced by a more complex one,
\begin{equation}                                             \label{Feff}
F_{\rm eff}\{n\} = \int d^2{\bf r}(f_2\{n\}+f_3\{n\} +f_4\{n\})\;,
\end{equation}
where the second term in the integrand,
\begin{equation}                                             \label{f3}
f_3\{n\}=-K_3[({\bf e}_1\nabla)n][({\bf e}_2\nabla)n][({\bf e}_3\nabla)n]\,,
\end{equation}
favors a finite slope in one of the three equivalent directions set by
the three unit vectors ${\bf e}_\alpha$ (where $\alpha=1,2,3$) forming
the angles of $120^{\,\circ}$ with each other.
The last term in the integrand, $f_4\{n\}$, is required to stabilize a finite value
of a slope. On general grounds, one can expect that the expansion of
$f_4\{n\}$ in powers of $\nabla n$ starts from the fourth-order contribution:
\begin{equation}                                              \label{f4}
f_4\{n\}=\frac{K_4}{4}(\nabla n)^4+\ldots\;.
\end{equation}

Note that the free energy functional (\ref{Feff}) has nothing in common
with the Ginzburg-Landau functional constructed by Domany {\em et al.}
\cite{DSWG}. In particular, in the latter the third-order term can appear
only in the absence of the particle-hole symmetry (which in terms of the spin
representation corresponds to 
$s_{\bf j}\Rightarrow -s_{\bf j}$).
In the model that we consider this symmetry is, naturally, always present, but
$F_{\rm eff}\{n\}$ 
has to contain the third-order term just as a consequence
of the symmetry of the problem in terms of the SOS representation.

From the form of  $F_{\rm eff}\{n\}$ it is clear that a phase
transition between the phases with zero and finite slopes cannot occur in
a continuous way.
The three equivalent auxiliary minima of $F_{\rm eff}\{n\}$ are formed at
finite values of $|\nabla n|$, and a first-order phase transition takes place
when the decrease of $K$ makes the free energy in these minima lower
than in the central minimum at $|\nabla n|=0$.

Thus we have obtained an additional confirmation of the conclusion that
a phase transition from the disordered phase with a zero slope to a phase
with a finite slope has to be of the first order.
Naturally, the construction of a phenomenological functional does not allow
one to distinguish whether it has to be a direct transition to the completely
frozen phase with the maximal possible slope,
or a transition to an intermediate phase with a smaller slope (that is,
with a finite concentration of spontaneously formed double domain walls),
which at lower temperatures will be followed by the second phase transition.

At finite $J_1$ the height variable $n$ can no longer be uniquely defined.
In terms of $n$ each plaquette at which the condition (\ref{cond}) is
violated corresponds to a screw dislocation on going around which $n$
changes by $\pm 6$ \cite{NHB}.
The core energy of such dislocations is close  to $2J_1$, whereas their
logarithmic interaction is too weak to keep them bound in pairs \cite{NHB}.
In such a situation the effective free energy should be a functional
not of a multivalued variable $n$, but of its derivatives,
\[
m_\alpha=({\bf e}_\alpha\nabla)n\,,
\]
which in the presence of free dislocations
do not have to satisfy the condition
\begin{equation}                                              \label{Summ}
m_1+m_2+m_3=0\,.
\end{equation}

This can be taken into account by making in Eqs.
\makebox{(\ref{f2})-(\ref{f4})} a replacement,
\[
(\nabla n)^2\Rightarrow \frac{2}{3}(m_1^2+m_2^2+m_3^2)\,,
~~~
({\bf e}_\alpha\nabla)n\Rightarrow m_\alpha\,,
\]
and adding to it a new contribution,
\[
f_{\rm D}=\frac{K_{\rm D}}{2}(m_1+m_2+m_3)^2
\]
(where $\ln K_{\rm D}\propto 2J_1/T$), which controls the fluctuations of
the difference between the densities of positive and negative dislocations.
However, the minimums of this new functional, which instead of the two
variables encoded in $\nabla n$ depends on the three variables $m_\alpha$, will
be achieved when they satisfy the condition (\ref{Summ}), and therefore
the conclusion on the first order of the transition
(obtained at $J_1=\infty$) will hold also at large but finite $J_1$.

\section{Comparison of different scenarios of disordering \label{md}}

In the two previous sections we have demonstrated that the phase transition
to the completely disordered phase (which can be associated with the
formation of a network of single domain walls) has to be of the first
order.
This still leaves possibilities for two different scenarios of disordering.
The formation of a network of domain walls can either happen at
$T=T_{\rm nw}>T_2$ as a separate phase transition, or at $T_{\rm nw}\leq T_2$.
In the latter case at $T>T_{\rm nw}$ the system is already in the
disordered phase (in which the domains of all six ground states are intermixed
with each other) and nothing special can be expected to happen at $T=T_2$.
In that case the only phase transition takes place at $T=T_{\rm nw}$.

In the Appendix we demonstrate how at $J_1=\infty$ one can construct
$T_c^+$, an exact upper boundary for $T_c$, the temperature of a phase
transition from the completely frozen phase, $T_c=\min\{T_2,T_{\rm nw}\}$.
The expression we obtain, Eq. (\ref{Tc+}), is applicable for an arbitrary
relation between $J_2$ and $J_3$.
In the case $J_3=0$ it gives \makebox{$T_c^{+}\approx 7.54\,J$,} which is
only 6\%
above the value of $T_2(J_3=0)=\gamma_2 T_2$ discussed in Sec. \ref{ddw}.
Since there are no reasons for this boundary from above
to give an extremely accurate estimate of $T_c$,
one can expect that the real value of the temperature at which
the fluctuations on the background of the completely frozen phase do
appear will be even lower than $T_2$, which means that for $J_3=0$
the single-transition scenario is realized.

Comparison of Eq. (\ref{Tddw}) with Eq. (\ref{Tnw}) shows that in the
region of parameters in which one can rather accurately estimate
both $T_2$ and $T_{\rm nw}$, these two temperatures coincide with each
other with the logarithmical accuracy.
This gives a hint that both mechanisms may be different manifestations of
the same phenomenon.

This idea looks even more plausible when one notices that the fluctuations of
a double domain wall considered in Sec. \ref{ddw} can also be discussed
in terms of zero modes. That is, the parallelogram in the middle
of Fig. \ref{3}(b) also can be considered as a domain which can be moved
along the direction of domain walls that are adjacent
to it without changing the total length of the walls.
Apparently, a typical network will have a less regular structure than the
network shown in Fig. \ref{4} and will incorporate some fragments looking
like finite pieces of a double wall. However, the contribution to
the free energy from each zero mode has to be of the form
\[
cE_{\rm dw}L+(E_{\rm a}+E_{\rm b})-T\ln L\,,
\]
where $c\sim 1$, which after summation over all modes will reproduce
the general structure of Eq. (\ref{Fnw}).
This suggests that there should be only one phase transition
which is of the first order and is related to the
appearance of a network formed both by single and double domain walls.

Comparison of Eq. (\ref{Tddw}) with Eq. (\ref{Tnw}) shows that
the realization of the two-transition scenario with $T_2<T_{\rm nw}$ requires
\makebox{$E_{\rm c}\ll E_{\rm mp}$}.
When only $J_1$, $J_2$ and $J_3$ are assumed to be non-zero, $E_{\rm c}$
satisfies the relation
\[
E_{\rm c}=E_a+E_b+E_{\rm dw}>E_{\rm mp}\,,
\]
and, therefore, any prerequisites for the phase transition splitting are
absent.
Nonetheless, they may appear when one takes into account the interaction
of more distant neighbors.

In particular, if one includes into
consideration the interaction of fifth neighbors (see Fig. \ref{1b}),
it turns out that 
$E_{\rm dw}$, $E_{\rm ddw}$ and $E_{\rm c}$ remain unchanged,
and therefore $T_2$ should not depend on $J_5$.
On the other hand, $E_{\rm mp}$ 
is increased by $8J_5$, which according to
Eq. (\ref{Tnw}) will shift the value of $T_{\rm nw}$ upwards.
Thus, one can expect that the increase of $J_5$ will lead
to the splitting of the phase transition into two.

However, the positiveness of $J_4$ works in the opposite direction and
therefore in a realistic system in which the coupling constants
continuously depend on the distance between sites, the influence of $J_5$
is likely to be compensated by the influence of $J_4$ (unless $J_4$ is,
for some reasons, of the opposite sign). Nonetheless, the possibility of phase
transition splitting is not entirely prohibited, although its realization
in some physical system is not very probable.

\section{Conclusion\label{conc}}

In the present work we have investigated the scenarios of disordering
of the striped phase which is formed
in a triangular-lattice Ising model with the antiferromagnetic interaction
of nearest and next-to-nearest neighbors.
Our analysis has shown that the destruction of such an ordering
has to take place via a single first-order phase transition.

The nature of this transition becomes more transparent in the case
$J_2-2J_3\ll J_2$, when it can be discussed in terms of the
formation of a {\em diluted} network of domain walls,
which is characterized by an extensive number of zero modes.
In this limit one can find how the transition temperature and the value of
the correlation radius at the transition point depend on the parameters of
the model [see Eq. (\ref{Tnw}) and Eq. (\ref{Lc})].
In the opposite limit of $J_3\rightarrow 0$
the transition temperature has to be proportional to $J_2$, the only energy
scale which is relevant for $J_2\ll J_1$, whereas the correlation radius
at the transition point has to be comparable with 1.

We also have shown that the formation of a domain-wall network could
be anticipated by a transition to the intermediate phase, in which
the long-range order in terms of $\sigma_{\bf j}$ is destroyed
by the spontaneous formation of a sequence of parallel (on the average)
double domain walls, whereas the long-range order in the orientation of
stripes formed by the spins of the same sign still exists.
This transition would be characterized by a combination of the Ising (in
the very narrow vicinity of the transition temperature) and the
Pokrovsky-Talapov (in a more wide temperature interval) critical behaviors.
However, in a system with only three coupling constants ($J_1$, $J_2$ and
$J_3$) no prerequisites for the realization of such a scenario  can be
found.

Nonetheless, they may appear when the interaction of more distant
neighbors is taken into account. In particular, the splitting of the phase
transition into two is favored by the increase of $J_5$. However, the
possibility of the realization of such a scenario depends on the fine interplay
between different coupling constants $J_k$ with $k\geq 4$ and in a
system with a monotonic dependence of $J_k$ on $k$ is not very probable.

An additional mechanism favoring the two-transition scenario may be
related to quantum fluctuations, whose role in decreasing
the energy of double walls may be more prominent than in decreasing
the energy of a diluted network due to a smaller size of moving objects
in the former case.

The numerical simulations of the triangular-lattice Ising antiferromagnet
have been performed in Refs. \makebox{\onlinecite{GP,NKL,RRT,TM}}.
In particular, Glosli and Plischke \cite{GP} have studied the system
with first and second neighbor interactions satisfying
$J_2/J_1=0.1$, Rastelli {\em et al.} \cite{RRT} - with
\makebox{$J_2/J_1=0.1,\,0.5,\,1$}, whereas Novikov {\em et al.} \cite{NKL}
have assumed that the interaction decays with the distance
exponentially. The results of these groups give evidence for the
existence of a single first-order transition,
which is consistent with our conclusions.
In the simulations of Takagi and Mekata \cite{TM} the disordering
of the striped phase has been investigated in the system with
\makebox{$J_2/J_1=0.2$} or $0.5$ and $J_3/J_1=-0.75$, but these authors make no
conclusions about the type of the single phase transition which they
observe.

In addition to more traditional applications mentioned in the Introduction,
the considered version of the Ising model can be used for the description
of a triangular array of quantum dots at half-filling \cite{NKL}
and (at sufficiently low temperatures) of a Josephson junction
array with the dice lattice geometry and one-third of the flux quanta
per plaquette \cite{xdt}.
In the latter case the role of $J_k$ with $k>1$ is played by the magnetic
interactions of currents in the array \cite{xdt,fxe}.
The results of this work may also be of help for understanding the nature
of phase transition(s) in the fully frustrated $XY$ model on a honeycomb
lattice \cite{ShS85,fxh}, in which the fluctuation-induced vortex pattern
\cite{fxh} has the same structure as in Fig. \ref{1a}.

This work has been supported in part by the Program ``Quantum
Macrophysics" of the Russian Academy of Sciences and
by the Program ``Scientific Schools of the Russian Federation"
(grant No.  1715.2003.2).

\appendix

\section{}

In the limit $J_1\rightarrow\infty$
it is convenient to split the Hamiltonian
(\ref{HIs}) into two terms, \[H=H_0+(H-H_0)\,,\] the first of which, $H_0$,
includes only the infinite interaction of nearest neighbors and restricts
the summation in the partition function to the states in which the
constraint (\ref{cond}) is satisfied on all triangular plaquettes,
whereas the second term, $H-H_0$, includes all other interactions.
It is well known \cite{Feyn} that the application of a variational procedure
allows one to use such a splitting to demonstrate that the free energy
of the system, $F$, is bounded from above by
\begin{equation}                                             \label{Fplus}
    F^{+}\equiv F_0+\langle H-H_0\rangle_0 \,,
\end{equation}
where $F_0$ is the free energy of the system whose Hamiltonian is equal to
$H_0$, whereas the angular brackets denote the average calculated with
the help of $H_0$.

In our case in the thermodynamic limit
\begin{eqnarray*} 
F_0 & = & -N(J_1+Ts_0)\,, \\
\langle H-H_0 \rangle_0 & = & 3N(J_2g_2+J_3g_3)\,,
\end{eqnarray*}
where $N$ is the total number of sites,
\[
    s_0\approx 0.323066
\]
is the residual entropy \cite{Wan}, whereas
$
g_{k}=\langle \sigma_{\bf i} \sigma_{\bf j}\rangle_0
$ 
is the correlation function of the variables ${\sigma}$ on the sites
${\bf i}$ and ${\bf j}$ which are the $k$th neighbors of each other,
calculated for the system with only nearest neighbor interaction at $T=0$.
According to Stephenson \cite{St64},
\[
g_2= \frac{1}{9}+\frac{2}{\sqrt{3}\pi}\,,~~~ 
g_3= \frac{1}{9}-\frac{3\;}{\pi^2}\,. 
\]

The comparison of $F^{+}(T)$
with the free energy of a completely frozen ground state,
which, naturally, coincides with its energy,
\[
E_0=-N(J_1+J_2-3J_3)\,,
\]
allows one to conclude that the temperature $T_c$, at
which a phase transition from a completely frozen state
to some phase with more developed fluctuations takes
place, cannot be larger than
\begin{equation}                                             \label{Tc+}
T_c^{+}(J_2,J_3)=C_2 J_2-C_3J_3\,,
\end{equation}
where
\begin{eqnarray}                                             \label{C2}
C_2 & = & {(1+3g_2)}/{s_0}\approx 7.54 \,,~
\\ 
C_3 & = & {3(1-g_3)}/{s_0}\approx 11.08\,.
\end{eqnarray}
Naturally, this approach does not allow to distinguish if the phase
transition at $T=T_c<T_c^+$ is a direct transition into the disordered
phase or a transition to the intermediate phase in the framework of
a two-transition scenario.

For $J_3=0$ the value of $T_c^+$ following from Eqs. (\ref{Tc+}) and (\ref{C2})
is rather close to $T_2$, the temperature of the spontaneous
formation of double domain walls given by Eq. (\ref{Tddw0}),
whereas in the limit of $E_{\rm dw}\rightarrow
0$ (that is \makebox{$J_3\rightarrow J_2/2$}) one gets
\[
T_c^+\approx 2.00\,J_2\,,
\]
which is compatible with an estimate for the temperature
of the phase transition given by Eqs. (\ref{Tddw}) and (\ref{Tnw}).



\begin{thebibliography}{99}
\bibitem{Ons}  
                L. Onsager, Phys. Rev. {\bf 65}, 117 (1944);
                G. H. Wannier,
                Rev. Mod. Phys. {\bf 17}, 50 (1945).

\bibitem{Wan}  G.H. Wannier, Phys. Rev. {\bf 79}, 357 (1950);
               Phys. Rev. B {\bf 7}, 5017E (1973).
\bibitem{Hout} R.M.F. Houtappel, Physica {\bf 16}, 425 (1950); 
               G.F. Newell, Phys. Rev. B {\bf 79}, 876 (1950); 
               K. Husimi and Y. Sy\^{ozi}, Prog. Theor. Phys. {\bf 5},
               177 and 341 (1950).
\bibitem{St64} J. Stephenson, J. Math. Phys. {\bf 5}, 1009 (1964).
\bibitem{St70} J. Stephenson, J. Math. Phys. {\bf 11}, 413 (1970).
\bibitem{BH}   H.W.J. Bl\"{o}te and H.J. Hilhorst,
               J. Phys. A {\bf 15}, L631 (1982).
\bibitem{Met}  B.D. Metcalf, Phys. Lett. {\bf 46A}, 325 (1974);
               M. Kaburagi and J. Kanamori,
               Jpn. J. Appl. Phys., Suppl.: pt. 2, 145 (1974).
\bibitem{NHB}  B. Nienhuis, H.J. Hilhorst and H.W.J. Bl\"{o}te,
               J. Phys. A {\bf 17}, 3559 (1984).
\bibitem{Weeks}J. D. Weeks, in {\em Order in Strongly Fluctuating
               Condensed Matter Systems}, edited by T. Riste
               (Plenum, New York - London, 1980), p. 293.
\bibitem{Land}D.P. Landau,    Phys. Rev. B {\bf 27}, 5604 (1983).
\bibitem{AP}    S. Alexander and P. Pincus,
                J. Phys. A {\bf 3}, 263 (1980).
\bibitem{FSK} S. Fujiki, K. Shutoh, and S. Katsura,
              J. Phys. Soc. Jpn. {\bf 53}, 1371 (1984).
\bibitem{MC}  S. Fujiki, K. Shutoh, Y. Abe and S. Katsura,
              J. Phys. Soc. Jpn. {\bf 52}, 1531 (1983);
              H. Takayama, K. Matsumoto, H. Kawahara and K. Wada,
              {\em ibid.}  {\bf 52}, 2588 (1983);
              S. Fujiki, K. Shutoh, S. Inawashiro, Y. Abe and S. Katsura,
              {\em ibid.}  {\bf 55}, 3326 (1986);
              S. Miyashita, H. Kitatani and Y. Kanada,
              {\em ibid.}  {\bf 60}, 1523 (1991);
              X. Qian and H.W.J. Bl\"{o}te,
              Phys. Rev. E {\bf 70}, 036112 (2004).
\bibitem{Mek} M. Mekata, J. Phys. Soc. Jpn. {\bf 42}, 76 (1977);
              M. Kaburagi, T. Tonegawa and J. Kanamori,
              {\em ibid.} {\bf 51}, 3857 (1982).
\bibitem{DSWG} E. Domany, M. Schick, J.S. Walker and R.B. Griffiths,
               Phys. Rev. B {\bf 18}, 2209 (1978).
\bibitem{KTK}  M. Kaburagi, T. Tonegawa and J. Kanamori,
               J. Magn. Magn. Mat. {\bf 31-34}, 1037 (1983).
\bibitem{SH}   P.A. Slotte and P.C. Hemmer,
               J. Phys. C {\bf 17}, 4645 (1984).
\bibitem{EH}   G. Einevoll and P.C. Hemmer,
               J. Phys. C {\bf 21}, L615 (1988).
\bibitem{DR}   E. Domany and E.K. Riedel,
               Phys. Rev. Lett. {\bf 40}, 561 (1978);
               Phys. Rev. B {\bf 19}, 5817 (1979).
\bibitem{KLF}  D. Kim, P.M. Levy and L.F. Uffer,
               Phys. Rev. B {\bf 12}, 989 (1975);
               A. Aharony, J. Phys. A {\bf 10}, 389 (1977).
\bibitem{NRS}  E.K. Riedel, Physica A {\bf 106}, 110 (1981);
               B. Nienhuis, E.K. Riedel and M. Schick,
               Phys. Rev. B {\bf 27}, 5625 (1983).
\bibitem{Sch}  M. Schick, Surf. Sci. {\bf 125}, 94 (1983).
\bibitem{OH}   S. Ostlund, Phys. Rev. B {\bf 24}, 398 (1981); 
               D.A. Huse, Phys. Rev. B {\bf 24}, 5180 (1981).
\bibitem{GP}   J. Glosli and M. Plischke,
               Can. J. Phys. {\bf 61}, 1515 (1983).
\bibitem{NKL}  D.S. Novikov, B. Kozinsky and L.S. Levitov,
               cond-mat/0111345 (2001).
\bibitem{RRT}  E. Rastelli, S. Regina and A. Tassi,
               Phys. Rev. B {\bf 71}, 174406 (2005).

\bibitem{TM}   T. Takagi and M. Mekata, {\bf 64}, 4609 (1995).
\bibitem{KK}   M. Kaburagi and J. Kanamori,
               J. Phys. Soc. Jpn. {\bf 44}, 718 (1978).
\bibitem{BMVW} P. Bak, D. Mukamel, J. Villain and K. Wentowska,
               Phys. Rev. B {\bf 19}, 1610 (1979).
\bibitem{PD}   R.E. Peierls,
               Proc. Cambridge Phil. Soc., {\bf 32}, 477 (1936);
               C. Domb, in {\em Phase Transitions and Critical Phenomena},
               vol. 3, ed. by C. Domb and M. S. Green
               (New York: Academic Press, 1974).
\bibitem{PT}   V.L. Pokrovsky and A.L. Talapov,
               Phys. Rev. Lett. {\bf 42}, 65 (1979);
               Zh. Eksp. Teor. Fiz. {\bf 78}, 269 (1980)
               [Sov. Phys. JETP {\bf 51}, 134 (1980)].
\bibitem{HF}   D.A. Huse and M.E. Fisher,
               Phys. Rev. Lett. {\bf 49}, 793 (1982);
               Phys. Rev. B {\bf 29}, 239 (1984).
\bibitem{Vil80a}J. Villain, in {\em Order in Strongly Fluctuating
               Condensed Matter Systems}, edited by T. Riste
               (Plenum, New York - London, 1980), p. 221.
\bibitem{Vil80b}J. Villain, Surf. Sci. {\bf 97}, 219 (1980); 
                J. Villain and M.B. Gordon, Surf. Sci. {\bf 125}, 1
                (1983).
\bibitem{CFHLB} S.N. Coppersmith, D.S. Fisher, B.I. Halperin,
                P.A. Lee and W.F. Brinkman,
                Phys. Rev. Lett. {\bf 46}, 549 and 869E (1981);
                Phys. Rev. B {\bf 25}, 349 (1982).
\bibitem{SW}   A.-C. Shi and M. Wortis,
               Phys. Rev. B {\bf 37}, 7793 (1988).
\bibitem{BPT}  T. Bohr, V.L. Pokrovsky and A.L. Talapov,
               Pis'ma ZhETF {\bf 35}, 165 (1982)
               [JETP Lett. {\bf 35}, 203 (1982)];
               T. Bohr, Phys. Rev. {\bf 25}, 6981 (1982).
\bibitem{xdt}  S.E. Korshunov, Phys. Rev. Lett. {\bf 94}, 087001 (2005).
\bibitem{fxe}  S.E. Korshunov, Phys. Rev. B {\bf 71}, 174501 (2005).
\bibitem{ShS85} W.Y. Shih and D. Stroud,
               Phys. Rev. B {\bf 32}, 158 (1985);
               S.E. Korshunov, J. Stat. Phys. {\bf 43}, 17 (1986).
\bibitem{fxh}  S.E. Korshunov and B. Dou\c{c}ot,
                Phys. Rev. Lett. {\bf 93}, 097003 (2004).
\bibitem{Feyn} R.P. Feynman, {\em Statistical Mechanics: A Set of Lectures}
               (Benjamen, Reading, 1972).

\end{thebibliography}
\end{document}